\documentclass[letterpaper, 10 pt, conference]{sty/ieeeconf}
\IEEEoverridecommandlockouts                       
\usepackage{amsmath,graphicx}

\def\isconfidential{1}
\def\includecomments{0}

\usepackage{sty/commands}
\usepackage{euscript}

\newcommand\Thetabf{\boldsymbol{\Theta}}
\newcommand\taubf{\boldsymbol{\tau}}
\newcommand\mubf{\boldsymbol{\mu}}
\newcommand\lambdabf{\boldsymbol{\lambda}}
\newcommand\alphabf{\boldsymbol{\alpha}}
\newcommand\betabf{\boldsymbol{\beta}}

\newcommand{\mg}[1]{\if\includecomments1\textcolor{green!50!black}{[MG:~#1]}\fi}

\title{\LARGE \bf
Dynamic Association of Semantics and Parameter Estimates by Filtering
}

\author{Marcus Greiff,
Ray Zhang,
Thomas Lew,
John Subosits
\thanks{$^{1}$All authors are with the Toyota Research Institute, 4440 El Camino Real, Los Altos, CA 94022. C.A.: 
        {\tt\small marcus.greiff@tri.global}.}%
}

\begin{document}

\maketitle
\thispagestyle{empty}
\pagestyle{empty}

\begin{abstract}
We propose a probabilistic semantic filtering framework in which parameters of a dynamical system are inferred and associated with a closed set of semantic classes in a map. We extend existing methods to a multi-parameter setting using a posterior that tightly couples semantics with the parameter likelihoods, and propose a filter to compute this posterior sequentially, subject to dynamics in the map's state. Using Bayesian moment matching, we show that the computational complexity of measurement updates scales linearly in the dimension of the parameter space. Finally, we demonstrate limitations of applying existing methods to a problem from the driving domain, and show that the proposed framework better captures time-varying parameter-to-semantic associations.
\end{abstract}

\section{Introduction}\label{sec:intro}
Recent work in legged locomotion has explored the possibility of improving visual segmentation by associating semantics from vision with parameters estimated through tactile sensing~\cite{ewen2022these,ewen2024you}. This approach entails an open association of estimated parameters to a closed set of semantic categories using the probabilistic Bayesian Moment Matching (BMM) framework developed in~\cite{jaini2016online}. This general framework has been proposed for driving applications~\cite{greiff2025semantic}, where the tactile sensing used in legged locomotion~\cite{ewen2024you} is replaced by estimates of vehicle parameters computed from on-board sensor measurements. The result is a \emph{semantic map} in which vehicle models are defined as a function of space. In contrast to~\cite{ewen2024you}, the objective in our work is not to improve segmentation, but rather to design a filter that better captures a latent map distribution for future use in predictive control and planning.

The semantic maps considered in our work are distinct from the occupancy-based maps in the robotics and control literature~\cite{borges2022survey,yang2017semantic,qian2024closing}. In the latter, points in the map are associated with occupancy~\cite{yang2017semantic}, class likelihoods~\cite{sunderhauf2016place}, traversability~\cite{borges2022survey}, or cost maps~\cite{goel2022predicting}. In contrast, the maps in~\cite{ewen2022these,ewen2024you,greiff2025semantic} associate points in the map with distributions over parameters of dynamical systems, permitting tighter dynamics-conditioned integration with model-based controllers~\cite{greiff2025semantic}.

While promising, the prior work~\cite{ewen2022these,ewen2024you,greiff2025semantic} assumes univariate parameter estimates. Furthermore, the association between the properties and semantic classes is assumed to be time-independent. These are both considerable limitations for driving applications, where physics-based models typically have several uncertain parameters, such as friction coefficients and cornering stiffness of the front and rear tires~\cite{thompson2024adaptive} or the weights of learned vehicle models~\cite{davydov2025first}. Furthermore, these parameters are generally time-varying and often estimated online. For example, friction coefficients change with tire wear and temperature~\cite{carlson2003nonlinear}. We generalize the method in~\cite{jaini2016online} used in~\cite{ewen2024you,greiff2025semantic} to address these limitations.

Instead of considering a static map, we model the map as a dynamical system, consider trajectories of the map's state in time, and implement the BMM scheme as a filter. At each time step, we output the marginal filtering posterior. This approach allows encoding dynamics in the map distributions' parameters, enabling similar exponential forgetting capabilities as in recursive least squares methods~\cite{aastrom1995adaptive}. 

\vspace*{-2.5pt}
\subsection{Contributions}\vspace*{-2.5pt}
We extend the framework of~\cite{ewen2024you,greiff2025semantic} in two ways. 
First, we generalize the normal-gamma representation to accommodate multivariate property measurements with shared semantic structure. 
Second, we formulate the semantic mapping as a filtering problem by introducing dynamics in the map's parameters. This induces exponential forgetting, capturing time-varying associations between properties and semantics.

\vspace*{-2.5pt}
\subsection{Notation}\label{sec:notation}\vspace*{-2.5pt}
Vectors are denoted by $\xvec=(x_1,...,x_n)\in \mathbb{R}^{n}$. The entry of a matrix $\XX$ on row $i$ and column $j$ of $\XX$ is denoted by $[\XX]_{ij}$. 
For a vector $\avec\in\Real^m$, we let $\mathrm{diag}(\avec)$ be a diagonal $m\times m$ matrix with $a_i$ on the $i$th diagonal element, and 0 elsewhere. We let $[K]=\mathbb{Z}_{[1,K]}$. The $K$-dimensional Dirichlet distribution is defined with a probability density function (PDF) $\mathrm{D}(\wvec|\avec) \propto \prod_{i=1}^K w_i^{a_i-1}$, and the categorical distribution is defined with a point mass function $\mathrm{C}(c|\wvec)\propto\prod_{i=1}^K w_i^{[c=i]}$, where $[\cdot]$ denotes the Iverson bracket, that is, $[c=i]=1$ if $c=i$ and 0 otherwise. The normal-gamma distribution is defined with a PDF $\mathrm{N}\Gamma(m, \tau|\mu, \lambda, \alpha, \beta)\propto \tau^{\alpha-\frac{1}{2}}\mathrm{exp}(-\beta\tau -\frac{\lambda\tau(m - \mu)^2}{2})$, and $\overline{\mathrm{N}\Gamma}(\mvec, \taubf|\mubf, \lambdabf, \alphabf, \betabf)=\prod_{i} \mathrm{N}\Gamma(m_i, \tau_i|\mu_i, \lambda_i, \alpha_i, \beta_i)$. $\mathbb{E}_{p(\xvec)}[\ubar\xvec]=\int\xvec p(\xvec)\der\xvec$ denotes the expectation of a random variable $\ubar{\xvec}$ sampled from a distribution $p(\xvec)$, and we sometimes omit $\ubar{(\cdot)}$ for clarity. $\psi(x)\hspace{-2pt}=\hspace{-2pt}(\der/\der x)\log(\Gamma(x))$ is the digamma function. We let $\Pcal$ denote a set of parameters and let its elements inherit annotations, i.e., if $\theta\in\mathcal{P}$, then $\bar{\theta}\in\bar\Pcal$, ${\theta}^t\in\Pcal^t$, and ${\theta}^{\star}\in\bar\Pcal^{\star}$.

\vspace*{-2.5pt}
\subsection{Outline}\vspace*{-2.5pt}
We first give necessary mathematical preliminaries in Sec.~\ref{sec:prelim}. The static mapping framework and BMM scheme are introduced in Sec.~\ref{sec:maps}, and extended to a filtering setting in Sec.~\ref{sec:dynamic}. Numerical examples are given in Sec.~\ref{sec:numerical}, highlighting the effect of using dynamic semantic-to-parameter associations for driving. Sec.~\ref{sec:conclusions} concludes the paper.

\section{Preliminaries}\label{sec:prelim}

In this work, we consider a probabilistic model of a parametric semantic map $p(\Thetabf)$, where $\Thetabf$ is a latent (state) variable. Given a measurement $\yvec$, we compute the posterior
\begin{equation}\label{eq:bayes_update}
p(\Thetabf|\yvec) = \frac{p(\yvec|\Thetabf)p(\Thetabf)}{p(\yvec)}.
\end{equation}
The computation of this posterior can be simplified by choosing an appropriate (conjugate) prior to the measurement likelihood. A prior $p(\Thetabf)$ belonging to a parametric family of distributions $\mathcal{Q}$ is said to be conjugate for the likelihood $p(\yvec|\Thetabf)$ if and only if the posterior $p(\Thetabf|\yvec)$ belongs to $\mathcal{Q}$. If the prior and likelihood are not conjugate, this approach suffers from an exponential increase in the number of parameters required to represent the posterior exactly. In such cases, we require a method of projecting the posterior onto $\mathcal{Q}$ to retain computational tractability, which can be done using Bayesian Moment Matching (BMM)~\cite{jaini2016online}. With BMM, we find the sufficient moments $g\in \mathbb{S}$ of the densities in $\mathcal{Q}$, and equate these with the moments of the posterior $p(\Thetabf|\yvec)$ in expectation. That is, for some new density $\bar{p}\in\mathcal{Q}$, we solve $\mathbb{E}_{\bar{p}(\Thetabf)}[g] = \mathbb{E}_{p(\Thetabf|\yvec)}[g]$ for the parameters of $\bar{p}$ to compute an approximate $\bar{p}(\Thetabf)\simeq p(\Thetabf|\yvec)$. While approximate, such BMM methods have proven useful in filtering~\cite{sarkka2023bayesian}, driving~\cite{greiff2025semantic}, and robotics~\cite{ewen2024you}. To use these ideas, we start by selecting a $p\in\mathcal{Q}$ and defining its sufficient moments.

\newcommand\cvec{\boldsymbol{c}}

\subsection{Sufficient Moments}
For the Dirichlet distribution, sufficient moments are $\{\log(w_i)\::\:i\in[K]\}$, and we have that $\mathbb{E}_{\mathrm{D}(\wvec|\avec)}[\log(w_i)] = \psi(a_i) - \psi(a_0)$ where $a_0 = \sum_i^Ka_i$. However, implementing BMM with this representation is impractical, as it would involve solving $\psi(a_i) - \psi(a_0) = c_i$ for all $i\in[K]$ for $\avec$ given constants $\cvec$. Instead, we follow~\cite{jaini2016online} and use sufficient moments for the Dirichlet distribution that are amenable to BMM (computationally) for a modest increase in memory.\begin{lemma}[Dirichlet, sufficient moments,~\cite{jaini2016online}]\label{lem:dirichletsuffstats}
A set of sufficient moments for the Dirichlet distribution $\mathrm{D}(\wvec|\avec)$ is $\mathbb{S}(\mathrm{D}) = \{(w_i, w_i^2)\;:\;i\in[K]\}$, with expectations
\begin{subequations}\label{eq:weightexpressionslem:dirichletsuffstats}
\begin{align}
&&&&\mathbb{E}[{w}_i]&=a_ia_0^{-1} &&\forall i\in [K],\\
&&&&\mathbb{E}[{w}_i^2]&=a_i(a_i+1)a_0^{-1}(a_0+1)^{-1} &&\forall i \in [K].
\end{align}
\end{subequations}
\end{lemma}

For the normal-gamma distribution, the natural moments involve logarithms and a digamma function in the expectation of these moments, similar to the Dirichlet distribution. Consequently, we choose an alternative set of moments.

\begin{lemma}[Normal-gamma, sufficient moments]\label{lemma:normalgamma:A}
\begin{subequations}A set of sufficient moments of the normal-gamma distribution $\mathrm{N}\Gamma(m,\hspace{-1pt} \tau;\hspace{-1pt} \mu,\hspace{-1pt} \lambda,\hspace{-1pt} \alpha,\hspace{-1pt} \beta)$ \hspace{-1pt}is\hspace{-1pt} $\mathbb{S}(\mathrm{N}\Gamma) \hspace{-2pt}=\hspace{-2pt} \{m, \hspace{-1pt}\tau, \hspace{-1pt}\tau^2, \hspace{-1pt}m\tau, \hspace{-1pt}m^2\tau\}$, with
\begin{align}
\mathbb{E}[m]&=\mu,\\
\mathbb{E}[{\tau}]&=\frac{\alpha}{\beta},\\
\mathbb{E}[\tau^2] &=\frac{\alpha + \alpha^2}{\beta^{2}},\\
\mathbb{E}[{m}{\tau}]&=\mu\frac{\alpha}{\beta},\\
\mathbb{E}[{m}^2{\tau}]&=\frac{1}{\lambda} + \mu^2\frac{\alpha}{\beta}.
\end{align}
\end{subequations}
\end{lemma}

\begin{proof}
The second expectation follows from the definition of variance. As the normal-gamma is an exponential distribution, the remaining expectations are obtained by taking the gradient of the log-normalizing constant when writing the PDF in its natural form.
\end{proof}

These two lemmas are instructive, as they both define a set of moments by which the Dirichlet and normal-gamma distributions can be recovered exactly, but also provide the expressions that can be used for BMM. Notably, if we construct intricate models composed as $p(\Thetabf) = \prod_ip_i(\Thetabf_i)$ with independent factors $p_i(\Thetabf_i)$, then $\mathbb{S}(p)=\bigcup_i \mathbb{S}(p_i)$. As a relevant example, consider a generalized normal-gamma distribution $\overline{\mathrm{N}\Gamma}(\mvec, \taubf| \mubf, \lambdabf, \alphabf,\betabf)$, which is the conjugate prior to a Gaussian likelihood $p(\yvec|\mvec, \taubf) = \mathrm{N}(\yvec|\mvec, \mathrm{diag}(\taubf)^{-1})$. We find that $\mathbb{S}(\overline{\mathrm{N}\Gamma}) = \{(m_j, \tau_j, \tau_j^2, m_j\tau_j, m_j^2\tau_j):j\in[J]\}$, and can therefore use this generalized normal-gamma distribution to construct the semantic map representation.

\section{Static Semantic Maps}\label{sec:maps}
We extend the BMM approach in \cite{greiff2025semantic,ewen2024you} to multivariate measurements $\yvec\in\Real^J$, where $J\geq 1$ is the dimension of each measurement. 
To derive an approximation to the posterior update~\eqref{eq:bayes_update}, we first define the parametric family $\mathcal{Q}$. Without loss of generality (see Appendix~\ref{app:extension}), we consider the problem of estimating the state of the map at an arbitrary point, which is henceforth omitted in the measurement likelihoods. We consider a single measurement, and extend this approach to a sequential filtering setting in Sec.~\ref{sec:dynamic}.

\subsection{Defining the prior}
Inspired by~\cite{ewen2024you}, we define the parametric family $\mathcal{Q}$ by a generalized Dirichlet-normal-gamma distribution with PDF
\begin{equation}
p(\Thetabf; \Pcal) = \mathrm{D}(\wvec|\avec)\prod_{i=1}^K\overline{\mathrm{N}\Gamma}(\mvec_{i}, \taubf_{i}| \mubf_{i}, \lambdabf_{i}, \alphabf_{i},\betabf_{i}),\label{eq:priormodel}
\end{equation}
where $K$ is the number of semantic classes, and
\begin{align}\label{eq:statespace}
\Thetabf &= \{(w_i, \mvec_i,\taubf_i)\in\Real_{\geq 0}\times\Real^J\times \Real^J_{>0}:i\in[K]\},\\
    \Pcal &=
    \{(a_i,\mubf_i, \lambdabf_{i}, \alphabf_{i},\betabf_{i})\in\Real_{>0}\hspace{-2pt}\times\hspace{-2pt} \Real^J\hspace{-2pt}\times\hspace{-2pt} \Real_{> 0}^{3J}\hspace{-1pt}:\hspace{-1pt}i\hspace{-1pt}\in\hspace{-1pt}[K]\},
\end{align}
where $\Thetabf$ is the latent state of the semantic property map and $\Pcal$ denotes its parameters. As all factors in~\eqref{eq:priormodel} are modeled as independent, we find its sufficient moments as
\begin{equation}\label{eq:suffmoments}
\hspace{-2pt}\mathbb{S}(p(\Thetabf; \hspace{-1pt}\Pcal)) \hspace{-2pt}=\hspace{-2pt}\{(w_i\hspace{-1pt},\hspace{-1pt}w_i^2\hspace{-1pt},\hspace{-1pt}\mvec_i\hspace{-1pt},\hspace{-1pt}\mvec_i\taubf_i\hspace{-1pt},\hspace{-1pt}\mvec_{i}^2\taubf_i\hspace{-1pt},\hspace{-1pt}\taubf_{i}\hspace{-1pt},\hspace{-1pt} \taubf_i^2)\hspace{-2pt}:\hspace{-2pt}i\hspace{-2pt}\in\hspace{-2pt}[K]\}\hspace{-1pt},\hspace{-2pt}
\end{equation}
where the products and squares are taken element-wise. The distribution proposed in~\eqref{eq:priormodel} is useful as it can be updated with both semantic information modeled as categorical likelihoods (see Sec~\ref{sec:catupdate}) and multivariate Gaussian mixtures, where the precision of each component is assumed to be unknown (see Sec.~\ref{sec:momentmatching}). Here, the independence assumptions enable a BMM scheme where computations scale linearly in $J$. Throughout this section, we assume that the map is static, in the sense that the true underlying map from which $\yvec$ is sampled can be expressed as in~\eqref{eq:priormodel} with time-invariant parameters. Later in Sec~\ref{sec:dynamic}, we relax this assumption to handle time-varying maps. Below, we provide a
short example in the driving domain.

\begin{figure}
    \centering
    \includegraphics[width=\columnwidth]{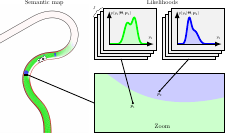}
    \vspace{-20pt}
    
    \caption{\emph{Semantic map in a driving application}. $K=3$ semantic classes (red, green, and blue) are inferred from images passed through a segmentation model, and the map is updated using a categorical likelihood~\eqref{eq:interpolatedcategorical}. The semantics are associated with properties through the Gaussian mixture model~\eqref{eq:extendedgmm}, here visualized along one of $J$ dimensions at two locations.}
    \label{fig:semantic-map}
\end{figure}

\begin{example}[Driving]\label{example:driving}
Consider the driving application in Fig.~\ref{fig:semantic-map}, analogous to that of~\cite{greiff2025semantic}. Semantic information on the road conditions is sampled (with $\ubar{c}\in[K]$) from an image taken by forward-facing cameras, which is passed through a segmentation model. In this example, colors correspond to one of three semantic classes, gravel ($\ubar{c}=1$, red), asphalt ($\ubar{c}=2$, green), and water ($\ubar{c}=3$, blue). Estimates of the vehicle's parameters are sampled from a parameter estimator (with $\ubar{\yvec}\in\Real^J$). These parameters could represent the friction coefficients of the front and rear wheels, respectively. The objective consists of updating the prior in~\eqref{eq:priormodel} with these measurements, thereby making the friction parameter distributions available (as a function of space) for use in a downstream control and planning stack. In Fig.~\ref{fig:semantic-map}, the parameter distributions are computed from the posterior at two points in space.
\end{example}

\subsection{Categorical Measurement Updates}\label{sec:catupdate}
To model semantic information in relation to~\eqref{eq:priormodel}, we consider a categorical measurement\hspace{-1pt} likelihood\hspace{-1pt} (see\hspace{-1pt} Sec.\hspace{-1pt}~\ref{sec:notation})
\begin{equation}\label{eq:interpolatedcategorical}
p(c|\Thetabf) = \mathrm{C}(c| \wvec).
\end{equation}
As the Dirichlet is the conjugate prior to the categorical likelihood~\cite{jaini2016online}, and this factor is modeled as independent of the normal-gammas in~\eqref{eq:priormodel}, we obtain a closed-form Bayesian update for the semantics similar to that described in~\cite{gan2020bayesian}.
\begin{lemma}[Posterior after categorical update]\label{lemma:posterior:form:cat}
Assuming a prior $p(\Thetabf; \Pcal)$ in~\eqref{eq:priormodel} and that $c$ is sampled from  a likelihood~\eqref{eq:interpolatedcategorical}, the posterior is $p(\Thetabf|c) = p(\Thetabf; \Pcal^{\star})$ where the normal-gamma parameters remain unchanged, and the new Dirichlet parameters of $\Pcal^{\star}$ are given by
\begin{align}
\begin{cases}
a^{\star}_i=a_i+1& \text{if}\;i=c,\\
a^{\star}_i=a_i & \text{otherwise}.
\end{cases}
\end{align}
\end{lemma}
\begin{proof}
Inserting~\eqref{eq:priormodel} and~\eqref{eq:interpolatedcategorical} in~\eqref{eq:bayes_update} yields the result.
\end{proof}

\subsection{Gaussian Mixture Measurement Updates}
\label{sec:momentmatching}
To model the multivariate properties, we assume that these can be measured directly as $\yvec$, and model the likelihood as a Gaussian mixture in the latent class probabilities 
\begin{equation}\label{eq:extendedgmm}
p(\yvec|\Thetabf) = \sum_{i=1}^Kw_i\mathrm{N}(\yvec| \mvec_i, \mathrm{diag}(\taubf_i)^{-1}).
\end{equation}
Unlike the categorical likelihood, there is no conjugate prior for this likelihood. However, given~\eqref{eq:priormodel}, the posterior computed by~\eqref{eq:bayes_update} is nonetheless known in closed form.

\begin{lemma}[Posterior after Gaussian mixture update]\label{lemma:posterior:form}
Assuming a prior~\eqref{eq:priormodel} and that $\yvec$ is sampled from  a measurement likelihood~\eqref{eq:extendedgmm}, the posterior distribution is
\begin{subequations}\label{eq:posterior}
\begin{align}
p(\Thetabf|&\yvec) = \frac{p(\yvec|\Thetabf)p(\Thetabf)}{p(\yvec)} \\
&= \frac{1}{M}\sum_{j=1}^K\Big(u_j\mathrm{D}(\wvec|\avec_j^\star)c_j^\star\overline{\mathrm{N}\Gamma}(\mvec_j, \taubf_j| \mubf_j^\star, \lambdabf_j^\star, \alphabf_j^\star,\betabf_j^\star)\notag\\
&\hspace{50pt}\prod_{i\neq j}\overline{\mathrm{N}\Gamma}(\mvec_i, \taubf_i | \mubf_i, \lambdabf_i, \alphabf_i,\betabf_i)\Big),
\end{align}
\end{subequations}
where,
\begin{subequations}\label{eq:posteriorparameterexpr}
\begin{align}
\hspace{-4pt}\mubf_j^\star &= \frac{\lambdabf_j\mubf_j + \yvec}{\lambdabf_j +1},&\hspace{-4pt}
\lambdabf_j^\star &= \lambdabf_j +\boldsymbol{1},\hspace{-4pt}\\
\hspace{-4pt}\alphabf_j^\star &= \alphabf_j + \tfrac12\boldsymbol{1},&\hspace{-4pt}
\betabf_j^\star &= \betabf_j + \lambdabf_j\frac{(\yvec - \mubf_j)^2}{2(\boldsymbol{1} + \lambdabf_j)},\hspace{-4pt}\\
\hspace{-4pt}\avec_j^\star &= \avec_j + \boldsymbol{1}_j, & \hspace{-4pt}
\hspace{-14pt}c_j^\star \hspace{-2pt}=\hspace{-2pt} 
\sqrt[J]{2\pi}&\mathrm{prod} \hspace{-2pt}\begin{pmatrix}\hspace*{-2pt}
\sqrt{\dfrac{\lambdabf_j}{\lambdabf_j^\star}}\dfrac{\Gamma(\alphabf_j^\star)}{\Gamma(\alphabf_j)}\dfrac{(\betabf_j)^{\alphabf_j}}{(\betabf_j^\star)^{\alphabf_j^\star}}\hspace*{-2pt}\end{pmatrix}\hspace*{-2pt},\label{eq:cstar}\hspace{-4pt}\\
\hspace{-4pt}M &= {\sum_{j= 1}^Ku_jc_j^\star}, &
\hspace{-4pt}u_j & =a_j \Big(\sum_{i=1}^K a_i\Big)^{-1}.\hspace{-4pt}
\end{align}
\end{subequations}
and all operations are done element-wise.
\end{lemma}

\begin{proof}
The result follows algebraically by completion of squares to match the normal-gamma factors, and leveraging the independence of these in the prior (see Appendix~\ref{app:posterior:form}).
\end{proof}

\subsection{Bayesian Moment Matching}
The number of terms in the posterior increases exponentially with the number of measurements, making it intractable for realistic implementations. However, this limitation can be remedied by BMM as outlined in Sec.~\ref{sec:prelim}. Given a set of sufficient moments for the generalized Dirichlet-normal-gamma model~\eqref{eq:priormodel} with parameters $\bar\Pcal$, we match the densities such that $p(\Thetabf; \bar\Pcal)\simeq p(\Thetabf|\yvec)$ in the sufficient moments. That is, we solve a system of equations $\mathbb{E}_{ p(\Thetabf; \bar\Pcal)}[g] =\mathbb{E}_{ p(\Thetabf|\yvec)}[g]$ for the parameters $\bar\Pcal$, expressed in the moments $\mathbb{E}_{ p(\Thetabf|\yvec)}[g]$.

\begin{lemma}[Moment-matched posterior]\label{lemma:BMMexpressions}
\begin{subequations}\label{eq:matched}Given a posterior~\eqref{eq:posterior}, the PDF $p(\Thetabf;\bar\Pcal)$ solving $\mathbb{E}_{ p(\Thetabf; \bar\Pcal)}[g] =\mathbb{E}_{ p(\Thetabf|\yvec)}[g]$ for $g\in\mathbb{S}(p(\Thetabf; \mathcal{P}))$ in~\eqref{eq:suffmoments} is parametrized by
\begin{align}
\bar\mubf_i &= \mathbb{E}[\mvec_i],\\
\bar\lambdabf_i &=\frac{1}{\mathbb{E}[\mvec_i^2\taubf_i]-\mathbb{E}[\mvec_i]^2\mathbb{E}[\taubf_i]},\\
\bar\alphabf_i &= \frac{\mathbb{E}[\taubf_i]^2}{\mathbb{E}[\taubf_i^2] - \mathbb{E}[\taubf_i]^2},\\
\bar\betabf_i &= \frac{\mathbb{E}[\taubf_i]}{\mathbb{E}[\taubf_i^2] - \mathbb{E}[\taubf_i]^2},\\
\bar a_i&= \mathbb{E}[w_i]\frac{\mathbb{E}[w_i]-\mathbb{E}[w_i^2]}{\mathbb{E}[w_i^2]-\mathbb{E}[w_i]^{2}},
\end{align}
\end{subequations}
for all $i\in[K]$, where operations are applied element-wise and the expectations in~\eqref{eq:matched} are taken with respect to~\eqref{eq:posterior}.
\end{lemma}
\begin{proof}
The result follows as all the factors of~\eqref{eq:posteriorparameterexpr} are modeled independently, see the details in the Appendix.
\end{proof}

To compute the posterior, we thus first evaluate its parameters~\eqref{eq:posteriorparameterexpr}, use them to express the moments $\mathbb{E}_{ p(\Thetabf|\yvec)}[g]$ with $g \in\mathbb{S}(p(\Thetabf; \mathcal{P}))$, and compute the matched posterior using Lemma~\ref{lemma:BMMexpressions}. In this particular case, the expectations in Lemma~\ref{lemma:BMMexpressions} can be expressed in closed form, as shown next. Specifically, let $g_i\in\mathbb{S}(p(\Thetabf ;\mathcal{P}))$ and define the function
\begin{align}
f(g_i) &= c_i^*u_i\mathbb{E}^\star[g_i] + \sum_{j\neq i}c_j^*u_j\mathbb{E}[g_i],\label{eq:ffunc}
\end{align}
where
\begin{align}
\mathbb{E}^\star[g_i] &= \int g_i \mathrm{D}(\wvec|\avec^{\star})\mathrm{N}\Gamma(\mvec_i, \taubf_i; \mubf_i^\star, \lambdabf_i^\star, \alphabf_i^\star,\betabf_i^\star)\der g_i,\notag\\
\mathbb{E}[g_i] &= \int g_i \mathrm{D}(\wvec|\avec)\mathrm{N}\Gamma(\mvec_i, \taubf_i; \mubf_i, \lambdabf_i^, \alphabf_i,\betabf_i)\der g_i,\notag
\end{align}
then $\mathbb{E}_{p(\Thetabf|\yvec)}[g_i] = f(g_i)/M$. \begin{subequations}\label{eq:exactmoments}For example, if $g_i=\mvec_i$, then
\begin{align}
\mathbb{E}_{p(\Thetabf|\yvec)}[\mvec_i] &= \frac{1}{M}f(\mvec_i)\\
&=\frac{1}{M}\Big(c_i^\star u_i\mathbb{E}^\star[\mvec_i] + \sum_{j\neq i}c_j^\star u_j\mathbb{E}[\mvec_i]\Big)\\
&= \frac{1}{M}\Big(c_i^\star u_i\mubf_i^\star + \sum_{j\neq i}c_j^\star u_j\mubf_i\Big).
\end{align}
\end{subequations}
The last equality follows by Lemma~\ref{lemma:normalgamma:A}, and can be repeated for all of the relevant moments in Lemma~\ref{lemma:BMMexpressions}, including the Dirichlet parameters, for which Lemma~\ref{lem:dirichletsuffstats} is used instead.

We can now update the prior~\eqref{eq:priormodel} with measurements drawn from the likelihoods in~\eqref{eq:extendedgmm} and~\eqref{eq:interpolatedcategorical}, just as in~\cite{ewen2024you,greiff2025semantic}. The only difference is that the proposed BMM scheme supports measurements where $J>1$. Next, we extend this framework further by considering time-varying map parameters.

\newcommand\timeindex{k}
\section{Dynamic Semantic Maps: A Filtering Problem}\label{sec:dynamic}
The general map representation in~\eqref{eq:priormodel} encodes an implicit assumption: that the association between the normal-gammas and the closed set of semantic classes is static. In essence, all measurements contribute to the posterior equally, and, disregarding the approximations done in the BMM, the order of processing and time between consecutive measurements does not matter. This assumption is not entirely realistic if the maps exhibit time variations in the property-to-class association. For example, suppose the semantic map is used to associate friction coefficients in a road-wheel model (properties) with certain types of road surfaces (classes). In this case, more recent measurements are likely to be more relevant than older measurements. 

To handle dynamic associations, we pose the semantic mapping as a filtering problem. Specifically, we let the sub-index $\timeindex$ denote a time-step, define a set of times $\{t_{\timeindex}\;:\;\timeindex\in\mathbb{N}\}$ with strictly positive and known time-steps between consecutive times $h_{\timeindex} = t_{\timeindex + 1} - t_{\timeindex} \geq 0$. We assume a prior $p(\Thetabf_0; \Pcal_{0|0})$ and use the notation $p(\Thetabf_{\timeindex}|\yvec_{0:\timeindex})= p(\Thetabf_{\timeindex}; \Pcal_{\timeindex|\timeindex})$, where $\Pcal_{a|b}$ denotes the parameters of the map up until $t_{a}$ conditioned on all measurements up until $t_{b}$, where $a\geq b$.

\subsection{Dynamics of the Map Parameters}

\begin{algorithm}[t!]
\caption{Dynamic BMM update of the semantic map.}
\label{alg:DSM}
\begin{algorithmic}[1]
\State Define parameter sets $\Pcal_{0|0}$, $\Pcal_{\infty}$, and a time constant $\Delta$.
\For{$\timeindex\in\mathbb{N}$}
\State \textbf{Receive:} $t_{\timeindex}, \yvec_{\timeindex}$
\AlgComment{Predict}
\State Set $h_{\timeindex}=t_{\timeindex}-t_{\timeindex-1}$
\State $\Pcal_{\timeindex|\timeindex-1}\leftarrow \mathrm{predict}(\Pcal_{\timeindex-1|\timeindex-1}, h_{\timeindex})$\hfill\eqref{eq:predmod}
\AlgComment{Update}
\State Eval. $\{(\avec_j^{\star}, \mubf_j^{\star},\lambdabf_j^{\star},\alphabf_j^{\star},\betabf_j^{\star})\}_{j}$ using $\yvec_{\timeindex},\Pcal_{\timeindex|\timeindex-1}$
\hfill\eqref{eq:posteriorparameterexpr}
\State Eval. $\{M, \cvec^{\star}, \uvec\}$ using $\{(\avec_j, \lambdabf_j^{\star},\alphabf_j^{\star},\betabf_j^{\star})\}_{j}$\hfill\eqref{eq:posteriorparameterexpr} 
\State Eval. $\mathbb{E}_{p(\Thetabf_{\timeindex}|\yvec_{0:\timeindex})}[g]$ for $g\in \mathbb{S}(p(\Thetabf_0))$ \hfill\eqref{eq:exactmoments}
\State $\Pcal_{\timeindex|\timeindex}\leftarrow \mathrm{project}(\Pcal_{\timeindex|\timeindex-1},\mathbb{E}_{p(\Thetabf_{\timeindex}|\yvec_{0:\timeindex})}[g])$ \hfill\eqref{eq:matched}
\EndFor
\end{algorithmic}
\end{algorithm}

We can now express a prediction model in these parameters, that is, how we expect them to evolve in the absence of measurement updates. To this end, we need to ensure that for any $\ubar{\Thetabf}_{\timeindex+1}\sim p(\Thetabf_{\timeindex+1}|\Thetabf_\timeindex)$, we have that $\ubar{\Thetabf}_{\timeindex+1}\in\mathbb{T}$ where $\mathbb{T}$ denotes the state-space of the map in~\eqref{eq:statespace}. To achieve this, we express a continuous ordinary differential equation (ODE) in the {parameters} of the map. We let $\mathbb{P}$ denote the parameter space in~\eqref{eq:statespace}, which involves a Cartesian product of real spaces and closed half-spaces, and steer a trajectory in this space towards a set of parameters $\Pcal_{\infty}\in\mathbb{P}$. Specifically, for times $t\in[t_{\timeindex}, t_{\timeindex+1})$, let
\begin{subequations}
\label{eq:parameterODE}
\begin{align}
({\der }/{\der t})\Pcal(t) &=  -\Delta^{-1}(\Pcal(t) - \Pcal_{\infty}),\\
\Pcal(t_{\timeindex})&=\Pcal_{\timeindex|\timeindex}\in\mathbb{P},
\end{align}
\end{subequations}
with a time-constant $\Delta>0$. Then $\Pcal(t_{\timeindex})\in\mathbb{P}\Rightarrow\Pcal(t)\in\mathbb{P}$ for all $t\in[t_{\timeindex},t_{\timeindex+1})$, $\lim_{t_{\timeindex+1}-t_{\timeindex}\to\infty}\Pcal(t_{\timeindex+1}) = \Pcal_{\infty}$, and the rate of convergence can be controlled by $\Delta$. Implementing the prediction model using~\eqref{eq:parameterODE} results in a weighting of past measurements, reducing their influence in time analogous to exponential forgetting in recursive least squares estimation schemes~\cite{aastrom1995adaptive}. In practice, we discretize~\eqref{eq:parameterODE} as
\begin{subequations}\label{eq:predmod}
\begin{align}
\Pcal_{\timeindex+1} &= c_{\timeindex}\Pcal_{\timeindex} + (1-c_{\timeindex})\Pcal_{\infty},\\
c_{\timeindex} &=\exp(-h_{\timeindex}\Delta^{-1}).\label{eq:constantzoh}
\end{align}
\end{subequations}
The algorithm is summarized in Alg.~\ref{alg:DSM}, and results in the semantic map of Sec.~\ref{sec:prelim} if $\c_{\timeindex}\equiv 1$ for all $t_{\timeindex}$ in~\eqref{eq:constantzoh}.

\begin{remark}
It would be possible to model dynamics in $\Thetabf_k$ directly. This approach would entail BMM of the Chapman-Kolmogorov equation~\cite{sarkka2013non}, which would in turn necessitate a Monte-Carlo approximation of the moments and incur significant computational cost. Modeling exponential forgetting in parameter space is intuitive and easy to tune in practice: $\Delta$ dictates how quickly the estimated map converges to the nominal map given by $\Pcal_{\infty}$ in the absence of measurements. 
\end{remark}

\begin{figure}[t!]
    \centering
    \includegraphics[width=\columnwidth]{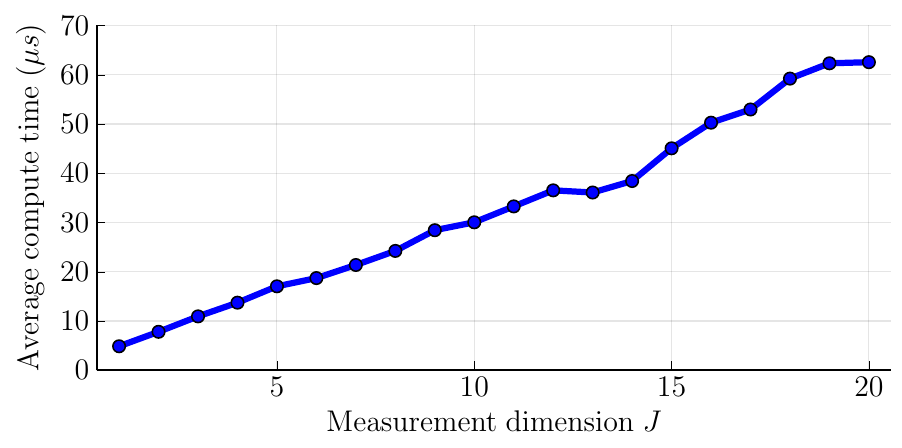}
    \vspace{-20pt}
    \caption{Average compute time over $10^6$ problems for a single Bayes update using the BMM scheme as a function of the measurement dimension $J$.}
    \label{fig:comptimes}
\end{figure}

\subsection{Computational Properties}

Compared with the univariate formulation in~\cite{ewen2024you}, the computation of the posterior parameters in~\eqref{eq:posteriorparameterexpr} and evaluation of the moments in~\eqref{eq:exactmoments} is approximately $J$ times more computationally expensive. The introduction of exponential forgetting through~\eqref {eq:predmod} has a negligible impact on compute time, shown as a function of $J$ in Fig.~\ref{fig:comptimes}. It increases linearly (in contrast to moment matching nonlinear filtering~\cite{berntorp2024framework}, which typically scales cubically in $J$) and remains in the order of <100$\mu$s for $J=20$ in an optimized Julia implementation. As such, it may be possible to build maps with larger parameter vectors, such as the linear last-layer meta-learned models in~\cite{lew2024risk}.

\begin{remark}\label{rem:extension}
In both the static and the dynamic cases, we can construct a semantic map using a set of voxels as in~\cite{ewen2024you}, where each is associated with a density~\eqref{eq:priormodel} while keeping the generalized normal-gamma component constant across the map (see Appendix~\ref{app:extension}). The only change to Alg.~\ref{alg:DSM} is then a step where the measurement $(t_{\timeindex},\yvec_{\timeindex})$ is associated with a specific voxel, and the update is done locally in that voxel.
\end{remark}

\section{Numerical Examples}\label{sec:numerical}
Next, we validate the proposed framework and demonstrate its efficiency. First, we consider a toy example where we can scale $(J,K)$ freely in Sec.~\ref{sec:examplemultivariate}. We then specialize this to a driving application described in Example~\ref{example:driving}, where the properties are parameter estimates of the tire model. We assume that there are different tires on the front and rear axles of a vehicle, as is common in vehicle racing~\cite{thompson2024adaptive,lew2024risk}. In Sec.~\ref{sec:timeinvariant}, we introduce time variation in the friction parameters, which become lower over time as the tires heat up and degrade~\cite{carlson2003nonlinear}. For simplicity and clarity of exposition, all experiments are performed in a single voxel of a larger map (see Remark~\ref{rem:extension}).

\begin{figure}[t!]
    \centering
    \includegraphics[width=\columnwidth]{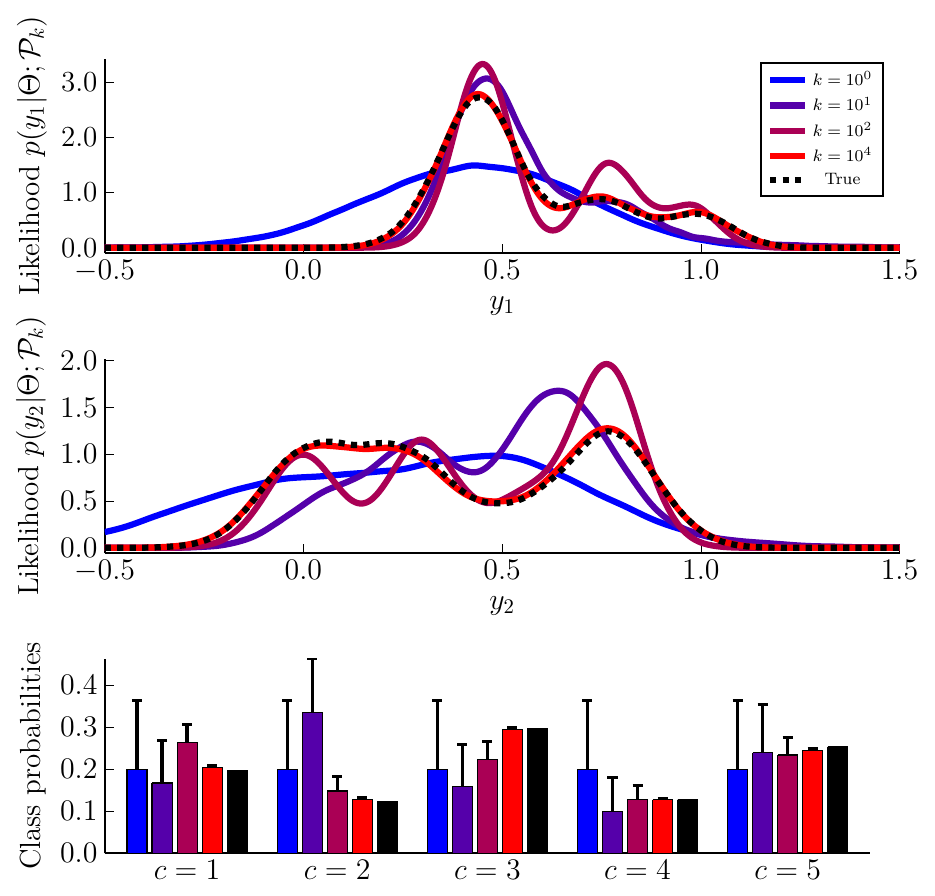}
    \vspace{-20pt}
    
    \caption{Convergence towards a true distribution (black) as a function of the number of samples $k$, from blue to red with $(J,K)=(5,5)$. Error bars indicate variance in class probabilities through the categorical likelihood, and only the first two dimensions of the property likelihood are shown.}
    \label{fig:multivariate}
\end{figure}

\newcommand\deltabf{\boldsymbol{\delta}}
\newcommand\truesymbol{t}
\subsection{Toy Example with Multivariate Properties}\label{sec:examplemultivariate}
First, we generate a target map with parameters $\Pcal^{\truesymbol}$ as
\begin{align*}
\avec^{\truesymbol} &\sim \Ucal([1,3]^K),&
\mubf^{\truesymbol}_i &\sim \Ucal([0,1]^J),&
\taubf^{\truesymbol}_i &\sim \Ucal([10^2,3\cdot 10^2]^J),\\
\lambdabf_i^{\truesymbol} &=\boldsymbol{1}\in\Real^J,&
\alphabf^{\truesymbol}_i &= 10^3\in\Real^J,&
\betabf^{\truesymbol}_i & = \alphabf_i^{\truesymbol} / \taubf_i^{\truesymbol}\in\Real^J.
\end{align*}
We also randomize a prior distribution with parameters $\Pcal_{0|0}$,
\begin{align*}
\mubf_i &=\mubf_{i,0|0} + \deltabf_i,&
(\lambdabf_{i,0|0},\alphabf_{i,0|0},\betabf_{i,0|0})  &=0.1(\lambdabf_i^{\truesymbol},\alphabf^{\truesymbol}_i,\betabf^{\truesymbol}_i),
\end{align*}
where $\deltabf_i\sim\Ucal([-0.5,0.5]^J)$ and $\avec_{0|0}=\boldsymbol{1}$. From this prior, we compute a posterior sequentially using Alg.~\ref{alg:DSM} in the static case where $\Delta\to \infty$ using only property measurements~\eqref{eq:extendedgmm}. The result is illustrated in Fig.~\ref{fig:multivariate}, where the prior converges to the true measurement likelihood as more measurements are included. It also demonstrates that the correct latent semantic representation emerges (class probabilities), despite not being explicitly measured through the categorical likelihood~\eqref{eq:interpolatedcategorical}.

\begin{figure}[t!]
    \centering
    \includegraphics[width=\columnwidth]{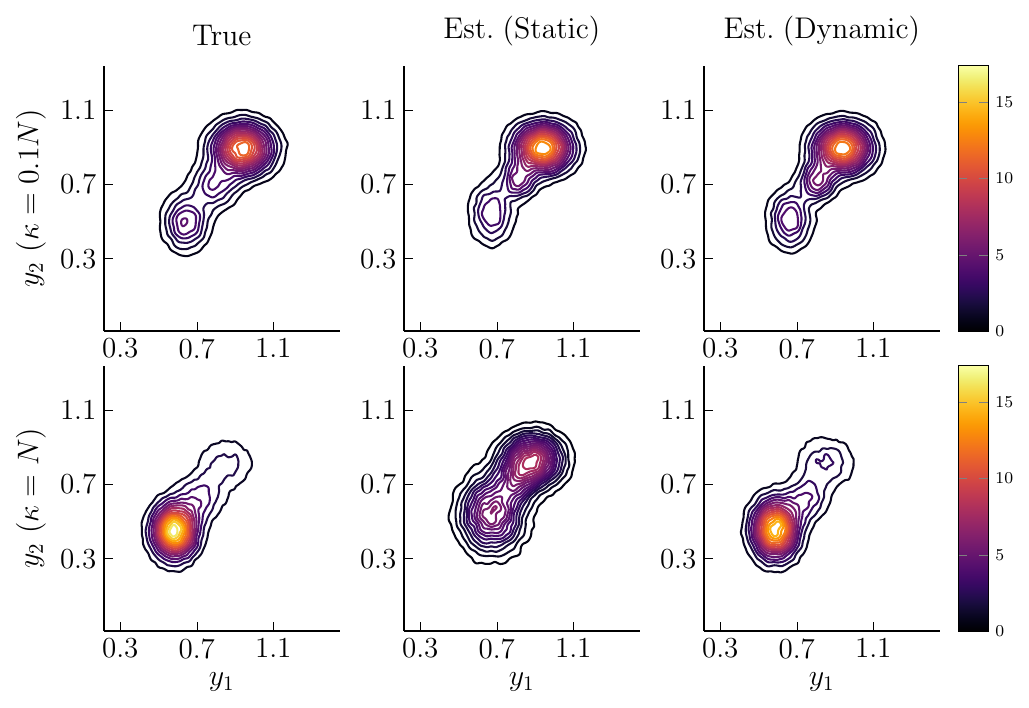}
    \vspace{-20pt}
    
    \caption{True and estimated likelihoods with the static BMM (center) and the dynamic BMM (right). The likelihoods are shown at two points in time, 10\% ($\timeindex = 0.1N$) and 100\% ($\timeindex = N$) into the simulation, respectively.}
    \label{fig:example2dlikelihoods}
\end{figure}

\subsection{Dynamic Friction Mapping in Driving Applications}\label{sec:timeinvariant}
Next, we consider Example~\ref{example:driving} with $J=2$ friction coefficients (front and rear tires, respectively) associated with $K=3$ different classes: dry asphalt ($c=1$), gravel ($c=2$), and wet asphalt ($c=3$). We now condition the voxel on property measurements sampled from~\eqref{eq:extendedgmm} given a true map associated with $\Pcal^t$ as well as semantic information sampled from~\eqref{eq:interpolatedcategorical}, both sampled at a rate of 10Hz. We consider a simulation from an initial time step ${\timeindex}=0$ to a terminal time step $\timeindex=N$. Additionally, we let $\Pcal^t$ vary in time by defining the true map with $\A_{\circ}=(50,10,10)$, $\A_{N}=(10,10,50)$,
\begin{align}
\hspace{-4pt}\M_{\circ} \hspace{-2pt}=\hspace{-2pt} 
\begin{bmatrix}
    0.95 & 0.8 & 0.65\\ 0.9 & 0.7 & 0.5
\end{bmatrix}\in\Real^{J\times K},\;\; \M_{N} \hspace{-2pt} =\hspace{-2pt}  0.9\M,\hspace{-4pt}
\end{align}
and, at the time step  $\timeindex$, letting
\begin{subequations}
\begin{align}
(\mubf_{1,\timeindex}^t,\mubf_{2,\timeindex}^t,\mubf_{3,\timeindex}^t) &= (1-\gamma_{\timeindex})\M_{\circ} + \gamma_{\timeindex}\M_N\\
\avec^t_{\timeindex} &=(1-\gamma_{\timeindex})\A_\circ + \gamma_{\timeindex}\A_N
\end{align}
\end{subequations}
with $\gamma_{\timeindex} = \timeindex/N$ and the remaining parameters sampled as in Sec.~\ref{sec:examplemultivariate}. The underlying semantics then change in time (e.g., due to rain), and the friction distributions associated with the semantic classes also change (e.g., due to tire wear and temperature). This setup allows studying the impact of forgetting in the BMM scheme. We consider two cases:
\begin{enumerate}
\item \textbf{Static}: The dynamics~\eqref{eq:constantzoh} defined with $\Delta\to\infty$, which in the case of $J=1$ corresponds to~\cite{ewen2024you}.
\item \textbf{Dynamic}: The dynamics~\eqref{eq:constantzoh} defined with $\Delta=50$, and a target distribution set arbitrarily as $\Pcal_{\infty}=\Pcal_{0|0}$. 
\end{enumerate}
We visualize the resulting predictive measurement likelihoods at $\timeindex = N/10$ and $\timeindex=N$ in Fig.~\ref{fig:example2dlikelihoods}, showing that both methods accurately represent the measurement likelihood initially. However, as the assumptions of static map parameters $\Pcal^t$ are increasingly violated, we end up with an averaging effect at $\timeindex=N$ in the static case without exponential forgetting. In contrast, introducing dynamics in the map parameters results in a correct measurement likelihood at $\timeindex = N$ (lower right). The proposed forgetting mechanism dramatically improves performance, and we caution that the BMM method in~\cite{jaini2016online} applied in~\cite{ewen2024you} may be brittle if there is underlying time variation in the parameters of the true map.

\begin{figure}[t!]
    \centering
    \includegraphics[width=\columnwidth]{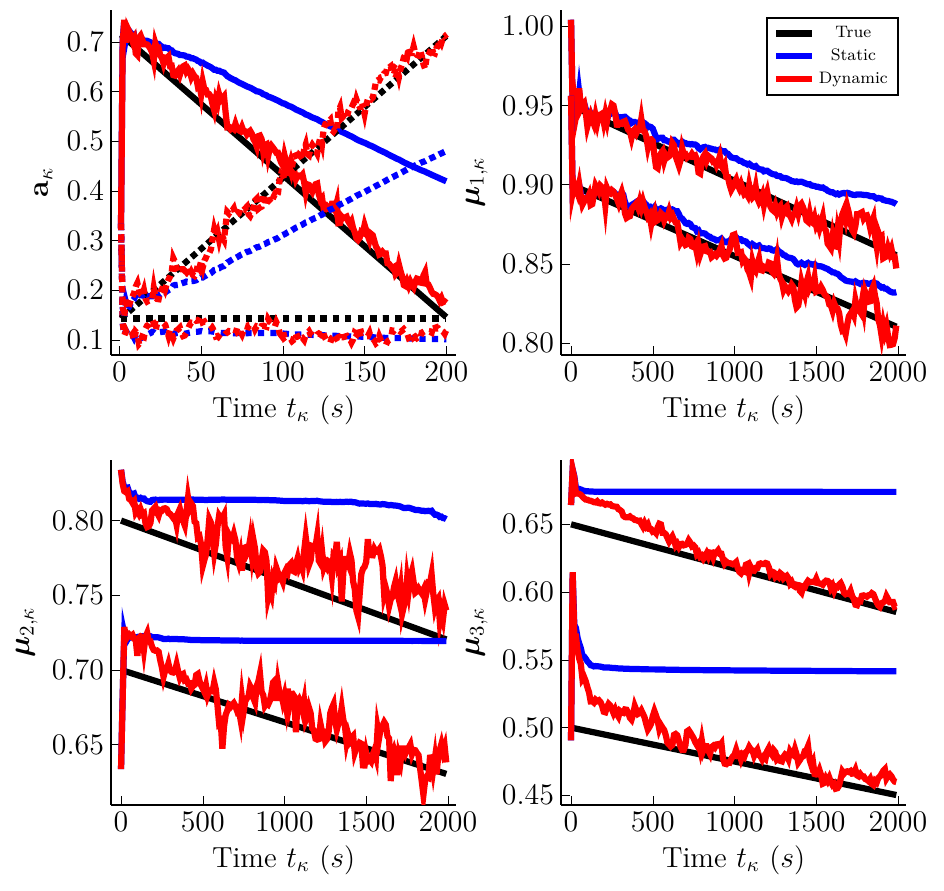}
    \vspace{-20pt}
    
    \caption{Trajectories of the latent parameters in the static case (blue) and dynamic case (red) versus the true map parameters (black) when the underlying parameters are time-varying $(\gamma = \timeindex/N)$.}
    \label{fig:exampleTV}
\end{figure}

To show the limitations of existing methods and consequences of violating the static map assumption, we present the time evolution of $(\avec_{\timeindex|\timeindex},\mubf_{1,\timeindex|\timeindex}, \mubf_{2,\timeindex|\timeindex}, \mubf_{3,\timeindex|\timeindex})_{\timeindex\in[N]}$ in Fig.~\ref{fig:exampleTV}. Introducing dynamics not only allows the BMM to capture the time variation in the latent semantic variables $\avec_{\timeindex}^t$, but also the more subtle changes in the latent normal gamma means. The effect of setting a low forgetting factor is that we get more noise in these estimates as a function of $\timeindex$, but faster adaptation. The signals that exhibit more noise (see e.g., $\mubf_{2,\timeindex|\timeindex}$) correspond to the class with the least probability, and are therefore unlikely to affect the predicted measurement likelihoods in Fig.~\ref{fig:example2dlikelihoods} significantly. However, a tradeoff needs to be made when choosing $\Delta$ in Alg.~\ref{alg:DSM}. 

\section{Conclusions}\label{sec:conclusions}
We have proposed a Bayesian moment-matching filtering framework for Dirichlet-normal-gamma distributions updated with samples from both categorical and multivariate Gaussian mixture likelihoods. We recover the mapping method in~\cite{ewen2024you} when the measurement dimension $J=1$ and the time constant $\Delta\to \infty$. Consequently, our approach extends the semantic mapping methods~\cite{ewen2024you} and~\cite{greiff2025semantic} to cases where $J>1$. In this setting, we show that the computational time of the Bayes update scales linearly in $J$. Additionally, we discuss how to define dynamics in the parameters of the Dirichlet-normal-gamma distribution. This approach yields exponential forgetting when used in a filtering scheme, making the estimator robust to slow time-varying changes in the state of the map. We have demonstrated improved performance for time-varying associates numerically in the context of a single point in the semantic map. Such maps are particularly relevant to driving applications where distributions of vehicle parameters, such as friction, can be associated with semantic classes, such as wet/dry road surfaces.

Future work will leverage the proposed semantic mapping framework for model predictive control in driving applications, with parameter estimators and segmentation models to sample from the measurement likelihoods. In particular, we will work on validating the independence and modeling assumptions on real data from the driving domain.

\appendix
We give two supporting lemmas before proving Lemma~\ref{lemma:posterior:form}, then give additional details on Lemma~\ref{lemma:BMMexpressions}, conclude with a discussion on extensions to spatial maps in Appendix~\ref{app:extension}.

\begin{lemma}[Normalization]\label{lem:direxpr}
Let $\mathrm{D}(\wvec|\avec)$ be a Dirichlet distribution, with $\wvec = (w_1,...,1_K)$ and $\avec=(a_1,...,a_K)$, then 
\begin{equation}\label{eq:direxpr}
w_j\mathrm{D}(\wvec|\avec) =  u_j\mathrm{D}(\wvec|\avec_j^{\star}),
\end{equation}
where $\avec_j^{\star} = (a_1,...,a_j+1,..., a_K)$ and $u_j =a_j/\sum_{j=1}^Ka_j$.
\end{lemma}
\begin{proof}
Let $a_0 = \sum_{j=1}^Ka_j$. We find that
\begin{align*}
w_j\mathrm{D}(\wvec|\avec) = \frac{\Gamma\begin{pmatrix}\sum\limits_{i=1}^Ka_i\end{pmatrix}}{\prod\limits_{i=1}^K\Gamma( a_i)}\prod\limits_{i=1}^K w_i^{a_i^{\star}-1}  = u_j \mathrm{D}(\wvec|\avec_j^{\star}),
\end{align*}
where the constant $u_j$ is given by
\begin{equation*}
u_j = \frac{\Gamma\begin{pmatrix}\sum\limits_{i=1}^Ka_i\end{pmatrix}}{\prod\limits_{i=1}^K\Gamma( a_i)}
\frac{\prod\limits_{i=1}^K\Gamma(a_i^{\star})}{\Gamma\begin{pmatrix}\sum\limits_{i=1}^K a_i^{\star}\end{pmatrix}} 
= \frac{\Gamma\begin{pmatrix}a_0\end{pmatrix}}{\Gamma\begin{pmatrix}1+a_0\end{pmatrix}}
\frac{\Gamma(a_j+1)}{\Gamma(a_j)}.
\end{equation*}
Since $\Gamma(x+1)=x\Gamma(x)$, we obtain
\begin{equation*}
u_j =\frac{\Gamma\begin{pmatrix}a_0\end{pmatrix}}{\Gamma\begin{pmatrix}1+a_0\end{pmatrix}}
\frac{\Gamma(a_j+1)}{\Gamma(a_j)} = \frac{\Gamma\begin{pmatrix}a_0\end{pmatrix}}{a_0\Gamma\begin{pmatrix}a_0\end{pmatrix}}
\frac{a_j\Gamma(a_j)}{\Gamma(a_j)}  = \frac{a_j}{a_0},
\end{equation*}
concluding the proof.
\end{proof}

The normal-gamma is the conjugate prior to the Gaussian measurement likelihood, but to implement BMM we require an explicit expression for the normalizing constant.
\begin{lemma}[Gaussian conjugate priors]\label{lem:gaussconj}
We have that
\begin{equation}
    \mathrm{N}(y; \hspace{-1pt}m, \hspace{-1pt}\tau^{-1}) \mathrm{N}\Gamma(m, \hspace{-1pt}\tau; \hspace{-1pt}\mu, \hspace{-1pt}\lambda, \hspace{-1pt}\alpha,\hspace{-1pt}\beta)\hspace{-1pt}=\hspace{-1pt} c^\star \mathrm{N}\Gamma(m,\hspace{-1pt}\tau; \mu^\star, \hspace{-1pt}\lambda^\star, \hspace{-1pt}\alpha^\star, \hspace{-1pt}\beta^\star)\label{eq:gammaidentity}
\end{equation}
\begin{subequations}where the parameters  are
\begin{align}
\mu^\star &=\frac{\lambda \mu + y}{(\lambda+1)}\\
\lambda^\star & =\lambda+1\\
\alpha^\star&=\alpha + \frac{1}{2}\\
\beta^\star &= \beta + \frac{\lambda (\mu - y)^2}{2(\lambda+1)}\\
c^\star &= \frac{1}{\sqrt{2\pi}}
\sqrt{\frac{\lambda}{\lambda^\star}}\frac{\Gamma(\alpha^\star)}{\Gamma(\alpha)}\frac{\beta^{\alpha}}{(\beta^\star)^{\alpha^\star}}.
\end{align}
\end{subequations}
\end{lemma}

\begin{proof} By definition and direct computations,
\begin{align}
&\mathrm{N}\Gamma(m, \tau; \mu, \lambda, \alpha,\beta)\mathrm{N}(y; m, \tau^{-1})\notag\\
=&\Big\{\frac{\sqrt{\lambda}}{\sqrt{2\pi}}\frac{\beta^{\alpha}}{\Gamma(\alpha)}\tau^{\alpha-\frac{1}{2}}\mathrm{exp}\Big(-\beta\tau -\frac{\lambda\tau(m - \mu)^2}{2}\Big)\Big\}\times\notag\\
&\Big\{\frac{1}{\sqrt{2\pi}}\tau^{\frac{1}{2}}\mathrm{exp}\Big(-\frac{\tau(y - m)^2}{2}\Big)\Big\}\notag\\
=&\frac{\sqrt{\lambda}}{2\pi}\frac{\beta^{\alpha}}{\Gamma(\alpha)}\tau^{\alpha}\mathrm{exp}\Big(-\beta\tau -\frac{\lambda\tau(m - \mu)^2}{2}-\frac{\tau(y - m)^2}{2}\Big)\label{eq:NIGA}.
\end{align}
By completion of squares
\begin{align*}
&M (m - \mu)^2 + Y (y - m)^2 =\\ &(M+Y)\Big(m - \frac{M\mu + Yy}{M + Y}\Big)^2 + \frac{YM(\mu - y)^2}{M + Y}.
\end{align*}
Letting $M = \frac{\lambda\tau }{ 2}$ and $Y = \frac{\tau }{ 2}$, we obtain
\begin{align*}
\eqref{eq:NIGA}
=&\frac{\sqrt{\lambda}}{2\pi}\frac{\beta^{\alpha}}{\Gamma(\alpha)}\tau^{\alpha+\frac{1}{2}-\frac{1}{2}}\mathrm{exp}\Big(\\
-&\underbrace{\Big(\beta + \frac{\lambda (\mu - y)^2}{2(\lambda+1)}\Big)}_{\beta^\star}\tau-\underbrace{\vphantom{\frac{\lambda (\mu - y)^2}{2(\lambda+1)})}(\lambda+1)}_{\lambda^\star}\frac{\tau}{2}\Big(m - \underbrace{\frac{\lambda \mu +    y}{(\lambda+1)}}_{\mu^\star}\Big)^2 \Big)\\
=&c^\star \mathrm{N}\Gamma(m,\tau; \mu^\star, \lambda^\star, \alpha^\star, \beta^\star).
\end{align*}
Finally, identifying $\{c^{\star}, \mu^\star, \lambda^\star, \alpha^\star, \beta^\star\}$ concludes the proof.
\end{proof}

\subsection{Proof of Lemma~\ref{lemma:posterior:form}}\label{app:posterior:form}
\begin{proof}Let $j\in[K]$ be a class index, $\mvec_j$ be a vector associated with class $j$, and $n\in[\mathrm{dim}(\yvec)]$ be a measurement dimension. In the following, the $n$th element of $\mvec_{j}$ is written $m_{jn}$. Assuming a prior~\eqref{eq:priormodel} and that $\yvec$ is sampled from  a measurement likelihood~\eqref{eq:extendedgmm}, the posterior distribution is
\begin{subequations}
\begin{align*}
p(\Thetabf|&\yvec) = \frac{p(\yvec|\Thetabf)p(\Thetabf)}{p(\yvec)} \\
&= \frac{1}{M}\sum_{j=1}^K\Big(w_j\mathrm{D}(\wvec|\avec_j)\mathrm{N}(\yvec; \mvec_j, \taubf_j^{-1})\\
&\hspace{50pt}\prod_{i=1}^K\overline{\mathrm{N}\Gamma}(\mvec_i, \taubf_i; \mubf_i, \lambdabf_i, \alphabf_i,\betabf_i)\Big)\\
&\underset{\eqref{eq:direxpr}}{=} \frac{1}{M}\sum_{j=1}^K u_j\mathrm{D}(\wvec|\avec_j^{\star})\prod_{i\neq j}\overline{\mathrm{N}\Gamma}(\mvec_i, \taubf_i; \mubf_i, \lambdabf_i, \alphabf_i,\betabf_i)\\
&\prod_{n=1}^{\dim(\yvec)}\mathrm{N}(y_n;m_{jn}, \tau_{jn}^{-1})\mathrm{N}\Gamma(m_{jn}, \tau_{jn}; m_{jn}, \lambda_{jn}, \alpha_{jn},\beta_{jn})\notag\\
&\underset{\eqref{eq:gammaidentity}}{=} \frac{1}{M}\sum_{j=1}^K u_j\mathrm{D}(\wvec|\avec_j^{\star})\prod_{i\neq j}\overline{\mathrm{N}\Gamma}(\mvec_i, \taubf_i; \mubf_i, \lambdabf_i, \alphabf_i,\betabf_i)\\
&\hspace{20pt}\prod_{n=1}^{\dim(\yvec)}c_{jn}^\star\mathrm{N}\Gamma(m_{jn}, \tau_{jn}; m_{jn}^\star, \lambda_{jn}^\star, \alpha_{jn}^\star,\beta_{jn}^\star)\notag\\
&= \frac{1}{M}\sum_{j=1}^K\Big(u_j\mathrm{D}(\wvec|\avec_j^\star)c_j^\star\overline{\mathrm{N}\Gamma}(\mvec_j, \taubf_j; \mubf_j^\star, \lambdabf_j^\star, \alphabf_j^\star,\betabf_j^\star)\notag\\
&\hspace{50pt}\prod_{i\neq j}\overline{\mathrm{N}\Gamma}(\mvec_i, \taubf_i; \mubf_i, \lambdabf_i, \alphabf_i,\betabf_i)\Big),
\end{align*}
\end{subequations}
where $\{c^{\star}_{jn}, \mu^\star_{jn}, \lambda^\star_{jn}, \alpha^\star_{jn}, \beta^\star_{jn}\}$ is computed as in Lemma~\ref{lem:gaussconj},  $\avec_{j}^{\star}$ as in Lemma~\ref{lem:direxpr}, and $c^{\star}_{jn} = \prod_{n=1}^{\mathrm{dim}(\yvec)}c^{\star}_{jn}$. We only require the normalizing constant $M$, which we can find by
\begin{align*}
&M = \int M p(\Thetabf|\yvec) \der\Thetabf
=\sum_{j=1}^Ku_j c_j^\star\\&\underbrace{\int\mathrm{D}(\wvec|\avec_j^\star)\der\wvec}_{=1}\underbrace{\int \overline{\mathrm{N}\Gamma}(\mvec_j, \taubf_j; \mubf_j^\star, \lambdabf_j^\star, \alphabf_j^\star,\betabf_j^\star)\der\mvec_j\der\taubf_j}_{=1}\notag\\
&\prod_{i\neq j}\underbrace{\int \overline{\mathrm{N}\Gamma}(\mvec_i, \taubf_i; \mubf_i, \lambdabf_i, \alphabf_i,\betabf_i)\der \mvec_i\der \taubf_i}_{=1}=\sum_{j=1}^Ku_j c_j^\star,
\end{align*}
which completes the proof.
\end{proof}

\subsection{Proof of Lemma~\ref{lemma:BMMexpressions}}
\begin{proof}
Let all expectations be taken with respect to $p(\Thetabf|\yvec)$. Clearly, from Lemma~\ref{lemma:normalgamma:A} and the independence assumption, 
$\bar \muvec_i=\mathbb{E}[\mvec_i]$ for all $i\in[K]$, and we compute
\begin{align*}
\begin{cases}
\bar\betabf_i\mathbb{E}[\taubf_i]   &= \bar\alphabf_i\\
\mathbb{E}[\taubf_i^2] &= \dfrac{\bar\alphabf_i + \bar\alphabf_i^2}{\bar\betabf_i^2}
\end{cases}& \Rightarrow \begin{cases}\bar\betabf_i = \dfrac{\mathbb{E}[\taubf_i]}{\mathbb{E}[\taubf_i^2] - \mathbb{E}[\taubf_i]^2}\\
\bar\alphabf_i = \dfrac{\mathbb{E}[\taubf_i]^2}{\mathbb{E}[\taubf_i^2] - \mathbb{E}[\taubf_i]^2}
\end{cases}\hspace{-6pt},
\end{align*}
and
\begin{align*}
\mathbb{E}[\mvec_i^2\taubf_i]= \frac{1}{\bar\lambdabf_i} + \bar\mubf_i^2\frac{\bar\alphabf_i}{\bar\betabf_i}
\Rightarrow \;\; \bar\lambdabf_i &= \Big(\frac{\bar\beta_i\mathbb{E}[\mvec_i^2\taubf_i]-\bar\muvec_i^2\bar\alphabf_i}{\bar\betabf_i}\Big)^{-1} \\
&=\frac{1}{\mathbb{E}[\mvec_i^2\taubf_i]-\mathbb{E}[\mvec_i]^2\mathbb{E}[\taubf_i]},
\end{align*}
Similarly, using the independence assumption and Lemma~\ref{lem:dirichletsuffstats},
\begin{align*}
\begin{cases}
\bar a_0 &\hspace*{-7pt}= \bar a_i \mathbb{E}[w_i]^{-1}\\
\bar a_0(\bar a_0+1)\mathbb{E}[w_i^2]  &\hspace*{-7pt}= \bar a_i(\bar a_i+1)
\end{cases}
\Rightarrow &
\bar a_i \hspace{-2pt} =\hspace{-2pt} \mathbb{E}[w_i]\frac{\mathbb{E}[w_i]-\mathbb{E}[w_i^2]}{\mathbb{E}[w_i^2]-\mathbb{E}[w_i]^{2}}
\end{align*}
We thus have closed-form expressions for the parameters of the moment-matched posterior $p(\Thetabf; \bar\Pcal)$, with $\bar\Pcal = \{\bar a_i, \bar\muvec_i, \bar\lambdabf_i, \bar\alphabf_i, \bar\betabf_i|i\in[K]\}$ expressed in the sufficient moments of the posterior $p(\Thetabf|\yvec)$, completing the proof.
\end{proof}

\subsection{Extension to Full Semantic Maps}\label{app:extension}
Due to the independence of the factors in~\eqref{eq:priormodel}, extensions to full semantic maps can be done along the lines of either~\cite{ewen2024you} or~\cite{greiff2025semantic}. In both cases, we model the map posterior in $L$ points $\pvec^{(\ell)}\in\Real^d$, with a complete map distribution $p(\Thetabf) = \prod_{\ell=1}^L\mathrm{D}(\wvec^{(\ell)}|\avec^{(\ell)})\prod_{i=1}^K\overline{\mathrm{N}\Gamma}(\mvec_{i}, \taubf_{i}| \mubf_{i}, \lambdabf_{i}, \alphabf_{i},\betabf_{i})$. In the context of ~\cite{ewen2024you}, each $\ell\in[L]$ is associated with a small volume of space $V^{(\ell)}\subset \Real^d$ (a voxel), and the posterior map distribution at a point $\pvec\in V^{(\ell)}$ is $\mathrm{D}(\wvec^{(\ell)}|\avec^{(\ell)})\prod_{i=1}^K\overline{\mathrm{N}\Gamma}(\mvec_{i}, \taubf_{i}| \mubf_{i}, \lambdabf_{i}, \alphabf_{i},\betabf_{i})$. Alternatively, if following~\cite{greiff2025semantic}, the likelihoods can be interpolated using an function $I: \Real^d\times \Real^d \mapsto [0,1]$ satisfying $\sum_{\ell=1}^LI(\pvec,\pvec^{(\ell)})=1$ for all $\pvec\in\Real^d$. In this case, we obtain a result similar to Lemma~\ref{lemma:posterior:form:cat} with respect to an interpolated categorical likelihood $p(c|\Thetabf, \pvec)\triangleq \prod_{\ell=1}^L C(\wvec^{(\ell)}|\avec^{(\ell)})^{I(\pvec,\pvec^{(\ell)})}$ and a result similar to Lemma~\ref{lemma:posterior:form} with respect to an interpolated Gaussian mixture likelihood $p(\yvec|\Thetabf, \pvec)\triangleq\sum_{\ell=1}^LI(\pvec,\pvec^{(\ell)})\sum_{i=1}^Kw_i\mathrm{N}(\yvec| \mvec_i, \mathrm{diag}(\taubf_i)^{-1})$. In this latter approach, we are not limited to a voxel-based map, but can use more generic graph-based representations of the map~\cite{greiff2025semantic}.

\bibliographystyle{IEEEbib}
\bibliography{references}

\end{document}